\documentclass[11pt]{article}

\usepackage{amsmath}

\begin{document}

\title{ E.M. Ovsiyuk\footnote{e.ovsiyuk@mail.ru},  V.V. Kisel, V.M. Red'kov\footnote{v.redkov@dragon.bas-net.by} \\
Particle with spin 1 in a magnetic field \\on the spherical plane
$S_{2}$ }

\date{}

\maketitle

\begin{abstract}
There are constructed exact solutions of the quantum-mechanical
equation for a spin $S=1$ particle in 2-dimensional Riemannian
space of constant positive  curvature,
 spherical  plane,  in presence of an external magnetic field,
 analogue of the homogeneous magnetic field in the Minkowski  space.
  A generalized formula for energy levels describing   quantization of the motion
  of the vector particle in magnetic field on the 2-dimensional space  $S_{2}$ has been found,
  nonrelativistic   and relativistic equations
  have been solved.

\end{abstract}

\section{Introduction}

The  quantization  of a quantum-mechanical particle in the
homogeneous magnetic field belongs to classical  problems in
physics  \cite{1,2,3, Landau-Lifshitz}. In 1985 -- 2010, a more
general problem in a curved Riemannian background,  hyperbolic and
spherical  planes,  was extensively studied \cite{Comtet-1985,
Comtet-1987,
 Aoki-1987, Groshe-1988,
Klauder-Onofri, Avron-Pnueli-1990, Plyushchay-1991(1),
Plyushchay-1991(2), Dunne-1992, Plyushchay-1995,
Alimohammadi-Shafei Deh Abad-1996, Alimohammadi-Mohseni
Sadjadi-1996, Onofri-2001, Negro et al-2001, Gamboa et al-2001,
Klishevich-Plyushchay-2002, Drukker et  al-2004,
Ghanmi-Intissar-2004, Correa-Jakubsky-Plyushchay-2009,
Alvarez-Cortes-Horvathy-Plyushchay-2009},
   providing us with a new system having intriguing
dynamics and symmetry, both on  classical and quantum levels.

Extension  to 3-dimensional hyperbolic and spherical spaces was
 performed recently.
In  \cite{4,5,6}, exact solutions for  a scalar particle in
extended problem, particle in  external magnetic field on the
background of Lobachevsky $H_{3}$  and Riemann $S_{3}$ spatial
geometries were found. A corresponding system in the frames of
classical mechanics was examined in
 \cite{7,8,9}. In the present paper, we consider a quantum-mechanical problem  a particle with spin $1/2$ described
 by the Dirac equation in  3-dimensional Lobachevsky  and Riemann space models
 in presence of the external magnetic field.

In the present paper, we will construct exact solutions for a
vector particle described by 10-dimensional  Duffin--Kemmer
equation in external magnetic field on the background of
2-dimensional spherical space  $S_{2}$.

10-dimensional Duffin--Kemmer equation for  a vector particle in a
curved space-time has the form \cite{Book-1}
$$
\{ \beta^{c} [\; i \hbar \; (\;  e_{(c)}^{\beta}
\partial_{\beta}  + {1\over 2} J^{ab} \gamma_{abc}  \; )\;
+\; {e \over c} A_{c}  \; ] \; - \;  mc \} \Psi  = 0 \; ,
\eqno(1.1)
$$

\noindent where   $\gamma_{abc}$ stands for Ricci rotation
coefficiemts,
 $A_{a} = e_{(a)}^{\beta}  A_{\beta} $ represent tetrad components of electromagnetic 4-vector  $A_{\beta} $;
 $J^{ab} =  \beta ^{a}\beta ^{b}- \beta^{b} \beta ^{a}$ are generators of 10-dimensional
represen\-tation of the Lorentz group. For shortness, we use
notation $ e/c \hbar \Longrightarrow e, \; mc/ \hbar
\Longrightarrow M$.

In the space  $S_{3}$ we will use the system of cylindric coordinates \cite{Olevsky}
$$
dS^{2} = c^{2} dt^{2} - \rho^{2}  [  \cos^{2} z ( d r^{2} +
\sin^{2} r \, d \phi^{2} ) + dz^{2} ],
$$
$$
 z \in [-\pi /2 , + \pi /2
], \; r \in [0, + \pi ] , \; \phi \in [0, 2 \pi ] . \eqno(1.2)
$$

Generalized expression for electromagnetic potential for an homogeneous magnetic field in  the curved model $S_{3}$
is given as follows
$$
 A_{\phi} = -2B \sin^{2} {r \over 2} = B\; ( \cos r -1 )\; .
\eqno(1.3)
$$

We will consider the above equation in presence of the field
in the model  $S_{3}$. Corresponding to cylindric coordinates  $x^{\alpha}= (t,r,\phi,z)$
a tetrad can be chosen as
$$
 e_{(a)}^{\beta}(x) = \left |
\begin{array}{llll}
1 & 0 & 0 & 0 \\
0 & \cos^{-1}z & 0 & 0 \\
0 & 0 & \cos^{-1}z\;\sin^{-1} r & 0 \\
0 & 0 & 0 & 1
\end{array} \right |  .
\eqno(1.4)
$$

\noindent
Eq.  (1.1) takes the form
$$
\left \{  i \beta^{0} { \partial \over \partial t} + {1 \over \cos z} \left (  i  \beta^{1}
{ \partial \over \partial r}        + \beta^{2}      { i  \partial _{ \phi} + e B
(\cos r -1) + i J^{12} \cos r  \over \sin r}  \right )   \right.
$$
$$
 \left.
  +
   i\beta^{3}  { \partial \over \partial  z}  +  i   {\sin z \over \cos z}  \; ( \beta^{1} J^{13}    +   \beta^{2} J^{23} )
          - M   \right \} \Psi  = 0 \; ,
\eqno(1.5)
$$

To separate the variables in eq. (1.5), we are to employ an explicit form
of the Duffin--Kemmer matrices $\beta^{a}$;  it will be most convenient to use so called cyclic representation
\cite{Book-5}, where  the generator  $J^{12}$ is of diagonal form (we specify matrices by blocks in accordance  with
( $1-3-3-3$)-splitting)
$$
\beta^{0} = \left | \begin{array}{rrrr}
 0       &   0        &  0  &  0 \\
 0  &  0       &  i  & 0  \\
  0  &   -i       &   0  & 0\\
   0  &  0       &   0  & 0
\end {array}
\right |, \qquad
\beta^{i} = \left |
\begin{array}{rrrrr}
  0       &  0       &    e_{i}  & 0       \\
    0   &  0       &   0      & \tau_{i} \\
   -e_{i}^{+}  &  0       &   0      & 0       \\
   0       &  -\tau_{i}&   0      & 0
\end {array} \right | \; ,
\eqno(1.6)
$$

\noindent where
 $e_{i}, \; e_{i}^{\;t}, \; \tau_{i}$ denote
 $$
e_{1} = {1 \over \sqrt{2}} ( -i, \; 0  , \; i )\; , \qquad e_{2} =
{1 \over \sqrt{2}} ( 1 , \; 0  , \;  1 )\; , \qquad e_{3} = ( 0 ,
i  , 0)\; , \;
$$
$$
\tau_{1} = {1 \over \sqrt{2}} \left |  \begin{array}{ccc} 0  &  1
&  0  \\ 1 &  0  &  1  \\ 0  &  1  &  0
\end{array} \right | , \qquad
\tau_{2}= {1 \over \sqrt{2}} \left |
\begin{array}{ccc} 0  &  -i  &  0  \\ i & 0  &  -i  \\ 0  &  i  &
0
\end{array} \right | , \qquad
  \tau_{3} =   \left |
\begin{array}{rrr} 1  &  0  &  0  \\ 0  &  0  &  0   \\ 0  &  0
&  -1
\end{array} \right |  =  s_{3}\; .
$$
$$
\eqno(1.7)
$$

\noindent The generator  $J^{12}$ explicitly reads
$$
 J^{12} =  \beta^{1} \beta^{2} -  \beta^{2} \beta^{1}
=  -i \left | \begin{array}{cccc}
0 & 0  &  0 & 0  \\
0 &   \tau_{3}  & 0 & 0 \\
0 & 0 &  \tau_{3} & 0 \\
0 & 0 & 0 &  \tau_{3}
\end{array} \right | = -iS_{3}.
\eqno(1.8)
$$

\section{Restriction to 2-dimensional model
 }

Let us restrict ourselves
to 2-dimensional case, spherical space  $S_{2}$ (formally it is sufficient in eq. (1.5)  to remove dependence on the variable
$z$  fixing its value by  $z=0$)
$$
\left [   i \beta^{0}  {\partial \over \partial t} + i   \beta^{1}
{\partial \over \partial r}        + \beta^{2}      { i \partial
_{ \phi} + e B (\cos r -1) + i J^{12} \cos r  \over  \sin r}
           -  M   \right ] \Psi  = 0 \; .
\eqno(2.1)
$$

 \noindent
With the use of substitution
$$
\Psi = e^{-i\epsilon t  }  e^{im\phi}   \left |
\begin{array}{c}
\Phi_{0}  (r) \\
\vec{\Phi}(r)  \\
\vec{E} (r)  \\
\vec{H} (r)
\end{array} \right | ,
\eqno(2.2)
$$

\noindent
eq.  (2.1)  assi=umes the form
 (introducing notation  $m + B (1 - \cos r)  = \nu (r)$)
$$
\left [  \epsilon  \;  \beta^{0}   +  i   \beta^{1}\; {\partial
\over  \partial r }        - \beta^{2}    \;   { \nu (r)   -  \cos
r \; S_{3}    \over  \sin r}       -  M   \right ] \left |
\begin{array}{c}
\Phi_{0}  (r) \\
\vec{\Phi} (r) \\
\vec{E} (r) \\
\vec{H} (r)
\end{array} \right | =0 \;  .
\eqno(2.3)
$$

After separation of the variables and using the notation
$$
{1 \over  \sqrt{2}}  ( {\partial  \over \partial r} +  { \nu -\cos r  \over  \sin r } ) = \hat{a}_{-} \;, \;\;
{1 \over  \sqrt{2}}  ( {\partial  \over \partial r} +  { \nu +\cos r  \over  \sin r } ) = \hat{a}_{+}\;, \;\;
{1 \over  \sqrt{2}}   ( {\partial  \over \partial r} +  { \nu   \over  \sin r } ) = \hat{a} \; ,
$$
$$
{1 \over  \sqrt{2}}  (- {\partial  \over \partial r} +  { \nu -\cos r  \over  \sin r } ) = \hat{b}_{-} \;, \;\;
{1 \over  \sqrt{2}}  (- {\partial  \over \partial r} +  { \nu +\cos r  \over  \sin r } ) = \hat{b}_{+}\;, \;\;
{1 \over  \sqrt{2}}   (- {\partial  \over \partial r} +  { \nu   \over  \sin r } ) = \hat{b}
$$

\noindent
we arrive at the radial system
$$
  -\hat{b}_{-}   E_{1}   - \hat{a}_{+}    E_{3}
 = M   \; \Phi_{0} \; ,
\qquad
   - i  \hat{b}_{-}    H_{1}    +  i \hat{a}_{+} \; H_{3}
    + i \epsilon  \;   E_{2}   = M  \;  \Phi_{2}\;,
    $$
    $$
        i  \hat{a}     H_{2} +i \epsilon  \;   E_{1}
  = M  \;   \Phi _{1}\;,
\qquad
  - i  \hat{b}  H_{2} +i \epsilon  \;   E_{3}
= M    \; \Phi_{3}\; ,
$$
$$
\eqno(2.4)
$$
$$
    \hat{a}  \Phi_{0}  - i  \epsilon  \;  \Phi_{1} = M \;     E_{1}\;,
\qquad
-  i  \hat{a}  \Phi_{2}  = M  \; H_{1}\;,
\qquad
   \hat{b}
\Phi_{0}  -i \epsilon  \;\Phi _{3}  = M    \; E_{3}\;,
$$$$
  i   \hat{b}   \Phi_{2}
   =  M  \;    H_{3}\;.
\qquad
-i \epsilon   \Phi_{2}    = M  \;  E_{2} \;,
\qquad
  i   \hat{b}_{-}   \Phi_{1}
 - i  \hat{a}_{+}  \Phi_{3}     = M  \;   H_{2} \; ,
$$
$$
\eqno(2.5)
$$

\section{ Nonrelativistic approximation}

Excluding  non-dynamical  variables
$\Phi_{0}, H_{1}, H_{2}, H_{3}$ with the help of equations
$$
 -\hat{b}_{-}   E_{1}   - \hat{a}_{+}    E_{3}      = M   \;  \Phi_{0} \; ,
\qquad
-  i  \; \hat{a}   \Phi_{2} = M  \; H_{1}\;,
$$
$$
 i   \hat{b}_{-}   \Phi_{1}  - i  \hat{a}_{+}  \Phi_{3}     = M  \;   H_{2} \;
 ,\qquad
  i   \; \hat{b}   \Phi_{2}    =  M  \;    H_{3}\;.
  \eqno(3.1)
  $$

\noindent  we get 6 equations (grouping them in pairs)
    $$
        i \hat{a}  \;
        ( i   \hat{b}_{-}   \Phi_{1}  - i  \hat{a}_{+}  \Phi_{3})
         + i \epsilon  \;   M E_{1}
  = M^{2}     \Phi _{1}\;,
$$
$$
     \hat{a} \;
      (-\hat{b}_{-}   E_{1}   - \hat{a}_{+}    E_{3}
  - i  \epsilon  \;  M\Phi_{1} = M ^{2}     e_{1}\;,
$$
$$
\eqno(3.2a)
$$
$$
   - i  \hat{b}_{-}  \; ( -  i   \; \hat{a} \;  \Phi_{2} )    +  i \hat{a}_{+}
(i   \; \hat{b} \;  \Phi_{2})
    + i \epsilon  \;   M E_{2}   = M ^{2}   \Phi_{2}\;,
    $$
$$
-i \epsilon \;   M \Phi_{2}       = M^{2}    E_{2} \;,
$$
$$
\eqno(3.2b)
$$
$$
  - i \; \hat{b}\;
  ( i   \hat{b}_{-}  \Phi_{1}  - i  \hat{a}_{+}  \Phi_{3})
+ i \epsilon  \;   M E_{3}  = M^{2}     \Phi_{3}\; ,
$$
$$
    \hat{b} \;
     (-\hat{b}_{-}   E_{1}   - \hat{a}_{+}    E_{3} )   -i \epsilon  \; M \Phi _{3}  = M  ^{2}  E_{3}\;,
$$
$$
\eqno(3.2c)
$$

Now we introduce big and small constituents
$$
\Phi_{1} = \Psi_{1} + \psi_{1}\; ,\qquad \Phi_{2} = \Psi_{2} + \psi_{2}\; ,\qquad \Phi_{3} = \Psi_{3} + \psi_{3}\; ,
$$
$$
i E_{1} =  \Psi_{1} - \psi_{1}\; , \qquad i E_{2} =  \Psi_{2} - \psi_{2}\; ,
 \qquad i E_{3} =  \Psi_{3} - \psi_{3}\; ,
$$

\noindent besides we should separate the rest energy by formal change $\epsilon \Longrightarrow \epsilon + M $;
summing and subtracting equation within each pair  (3.2) and ignoring small
constituents
 $\psi_{i}$ we arrive at three equations for big  components
$$
\left ( - 2 \; \hat{a} \hat{b}_{-}      + 2\epsilon M  \right ) \psi_{1}    = 0 \; ,
$$
$$
\left ( -   \hat{b}_{-} \hat{a}  - \hat{a}_{+} \hat{b} +2\epsilon
M     \right )  \psi_{2}
  =0 \; ,
$$
$$
\left ( -  2  \hat{b} \hat{a}_{+}
  +2\epsilon M \right )  \psi_{3}    = 0 \; .
  \eqno(3.3)
$$

\noindent
It is a needed Pauli-like system  for the spin 1 particle.

Allowing for  $\nu (r) =m+B\,(1-\cos r)$, from (3.3) we get radial equations in the form
$$
\left[{d^{2}\over dr^{2}}+{\cos r\over \sin r}\,{d\over dr}-B-
{1-2\, [ m+B\,(1-\cos r) ]\,\cos r\over \sin^{2}r}-
\right.
$$
$$
\left. - {[m+B\,(1-\cos r)]^{2}\over \sin^{2}r}+2\,\epsilon\, M \right]\psi_{1}=0\,,
$$
$$
\eqno(3.4a)
$$
$$
\left[{d^{2}\over dr^{2}}+{\cos r\over \sin r}\,{d\over dr}-
{[ m+B\,(1-\cos r)]^{2}\over \sin^{2}r}+2\,\epsilon \, M\right]\psi_{2}=0\,,
$$
$$
\eqno(3.4b)
$$
$$
\left[{d^{2}\over dr^{2}}+{\cos r\over \sin r}\,{d\over dr}+B-{1+2\,[m +B\,(1-\cos r)] \,\cos r\over \sin^{2}r}-
\right.
$$
$$
\left. -
{[m+B\,(1-\cos r)] ^{2}\over \sin^{2}r}+2\,\epsilon\, M \right]\psi_{3}=0\,.
$$
$$
\eqno(3.4c)
$$

The first and the third equations are symmetric with respect to  formal change
$
m \Longrightarrow -m\,, \;B \Longrightarrow - B$.

Let us  consider eq.  $(3.4a)$. In the new variable
$
1-\cos r =2 \,y\,;
$ and with the use of a substitution
$
B_{1} = y^{C_{1}} (1-y) ^{A_{1}}  f_{1}
$ the differential equation assumes the form
$$
y\,(1-y)\,{d^{2}B_{1}\over dy^{2}}+[2\,C_{1}+1-(2\,A_{1}+2\,C_{1}+2)\,y]\,{dB_{1}\over dy}
$$
$$
+\left[B^{2}+B+2\,\epsilon\,M-(A_{1}+C_{1})\,(A_{1}+C_{1}+1)\right.
$$
$$\left.+\,
{1\over 4}\,{4\,A_{1}^{2}-(2\,B+m+1)^{2}\over1-y}+{1\over 4}\,{4\,C_{1}^{2}-(m-1)^{2}\over y}\right]B_{1}=0\,.
\eqno(3.5)
$$

At  $A_{1},\,C_{1}$ taken according to
$$
A_{1}= + {1\over 2} \mid 2B+m+1 \mid \; ,\qquad C_{1}= + {1\over 2}
\mid m-1 \mid \,,\eqno(3.6)
$$

\noindent eq.  $(3.5)$ is recognized as a hypergeometric equation \cite{Bateman} with parameters
$$
\alpha_{1}=A_{1}+C_{1}+{1\over 2}-\sqrt{B^{2}+B+2\,\epsilon\,M+{1\over 4}}\,,
$$
$$
\beta_{1}=A_{1}+C_{1}+{1\over 2}+\sqrt{B^{2}+B+2\,\epsilon\,M+{1\over 4}}\,,
$$
$$
\gamma_{1}=2\,C_{1}+1\,,\qquad B_{1}=y^{C_{1}}\,(1-y)^{A_{1}}\,F\,(\alpha_{1},\,\beta_{1},\,\gamma_{1};\;y).
\eqno(3.7)
$$

To get polynomials we need to impose restriction
 $\alpha_{1}=-n$,  from this it follows
$$
\psi_{1}=y^{C_{1}}(1-y)^{A_{1}}\,F(\alpha_{1},\beta_{1},\gamma_{1};\,y)\,,
$$$$
\sqrt{B^{2}+B + 2\,\epsilon\,M+{1\over 4}} = n + {1 \over 2} +  { \mid 2B + m +1  \mid +  \mid m-1 \mid \over 2} \; .
\eqno(3.8)
$$

In similar manner,  we construct solution of eq. (3.4b)
$$
\psi_{2}=y^{C_{2}}(1-y)^{A_{2}}\,F(\alpha_{2},\beta_{2},\gamma_{2};\,y)\,,
$$
$$
 A_{2}=\pm\,{1\over 2}\,(2\,B+m)\,,\;\; C_{2}= \pm\,{m\over 2}\,,\;\;\gamma_{2}=2\,C_{2}+1\,,
$$
$$
\alpha_{2}=A_{2}+C_{2}+{1\over 2}-\sqrt{B^{2}+2\,\epsilon\,M+{1\over 4}}\,,
$$
$$
\beta_{2}=A_{2}+C_{2}+{1\over 2}+\sqrt{B^{2}+2\,\epsilon\,M+{1\over 4}}\,;
\eqno(3.9)
$$

\noindent from quantization  condition $\alpha_{2}=-n$ it follows
$$
\sqrt{B^{2}+2\,\epsilon\,M+{1\over 4}} = n + {1 \over 2} +  { \mid 2 B + m  \mid +  \mid m \mid \over 2} \; .
\eqno(3.10)
$$

Finally, we construct solutions for eq.  $(3.4c)$:
$$
\psi_{3}=y^{C_{3}}(1-y)^{A_{3}}\,F(\alpha_{3},\beta_{3},\gamma_{3};\,y)\,,
$$
$$
A_{3}=\pm\,{1\over 2}\,(2\,B+m-1)\,,\;\; C_{3}= \pm\,{1\over 2}\,(m+1)\,,\;\;\gamma_{3}=2\,C_{3}+1\,,
$$
$$
\alpha_{3}=A_{3}+C_{3}+{1\over 2}-\sqrt{B^{2}-B+2\,\epsilon\,M+{1\over 4}}\,,
$$
$$
\beta_{3}=A_{3}+C_{3}+{1\over 2}+\sqrt{B^{2}-B+2\,\epsilon\,M+{1\over 4}}\,;
\eqno(3.11)
$$
and requiring $\alpha_{3}=-n$ we obtain
$$
\sqrt{B^{2}- B + 2\,\epsilon\,M+{1\over 4}} = n + {1 \over 2} +  { \mid 2B + m -1  \mid +  \mid m+1 \mid \over 2} \; .
\eqno(3.12)
$$

\section{Solution of radial equations in relativistic case
}

Let start with eqs.  (2.4)--(2.5). Excluding six components  $E_{i}, H_{i}$ with the  help of (2.5), we derive
four second order equations for  $\Phi _{a}$:
$$
(- \hat{b}_{-} \hat{a} - \hat{a}_{+} \hat{b}  + \epsilon^{2} - M^{2} ) \Phi_{2} = 0 \; ,
$$
$$
(- \hat{b}_{-} \hat{a} - \hat{a}_{+} \hat{b}
 - M^{2} ) \Phi_{0} + i \epsilon ( \hat{b}_{-} \Phi_{1} + \hat{a}_{+}\Phi_{3})=0\; ,
$$
$$
(- \hat{a} \hat{b}_{-} + \epsilon^{2} - M^{2}) \Phi_{1} +
\hat{a} \hat{a}_{+} \Phi_{3} + i \epsilon \hat{a} \Phi_{0} = 0 \; ,
$$
$$
(- \hat{b} \hat{a}_{+} + \epsilon^{2} - M^{2}) \Phi_{3} +
\hat{b} \hat{b}_{-} \Phi_{1} + i \epsilon \hat{b} \Phi_{0} = 0 \; .
\eqno(4.1)
$$

Once, it should be noted existence of a simple solution of the system
$$
\Phi_{0} = 0\;, \qquad \Phi_{1}=0\;, \qquad \Phi_{3}=0 \; ,
\qquad
(- \hat{b}_{-} \hat{a} - \hat{a}_{+} \hat{b}  + \epsilon^{2} - M^{2} ) \Phi_{2} = 0 \; ,
\eqno(4.2a)
$$

\noindent at this from (2.5) it follows
$$
E_{1} = 0 \;, \qquad    E_{2} = -i \epsilon  M^{-1}  \Phi_{2}      \; , \qquad  E_{3} =0 \; ,
$$
$$
H_{1} = -  i M^{-1}  \hat{a} \; \Phi_{2}  \;,\qquad     H_{2} = 0 \;,\qquad     H_{3}= i  M^{-1} \hat{b}  \; \Phi_{2}
    \; .
\eqno(4.2b)
$$

Lets us turn to (4.1) and act on  the third equation from the left by operator $\hat{b}_{-}$,
and on the forth  equation by operator $\hat{a}_{+}$. Thus, introducing the notation
$$
\hat{b}_{-} \Phi_{1}  = Z_{1} \; , \qquad \hat{a}_{+} \Phi_{3} = Z_{3} \;,
$$

\noindent instead of (4.1)  we obtain
$$
(- \hat{b}_{-} \hat{a} - \hat{a}_{+} \hat{b}  + \epsilon^{2} - M^{2} ) \Phi_{2} = 0 \; ,
$$
$$
(- \hat{b}_{-} \hat{a} - \hat{a}_{+} \hat{b}
 - M^{2} ) \Phi_{0} + i \epsilon ( Z_{1} + Z_{3} )=0\; ,
$$
$$
(- \hat{b}_{-} \hat{a}  + \epsilon^{2} - M^{2})  Z_{1}  +
\hat{b}_{-} \hat{a} Z_{3} + i \epsilon \hat{b}_{-} \hat{a} \Phi_{0} = 0 \; ,
$$
$$
(-  \hat{a}_{+} \hat{b} + \epsilon^{2} - M^{2}) Z_{3}  +
\hat{a}_{+} \hat{b} Z_{1}  + i \epsilon  \hat{a}_{+} \hat{b} \Phi_{0} = 0 \; .
\eqno(4.3)
$$

\noindent
Instead of $Z_{1}, Z_{3}$, let us use new  variables
$$
Z_{1} ={f + g \over  2} \; , \qquad  Z_{3} ={f - g \over  2} \; ,
$$
$$Z_{1} +Z_{3} = f \;, \qquad Z_{1} - Z_{3} = g \; ;
$$

\noindent
then the system assumes the form
$$
(- \hat{b}_{-} \hat{a} - \hat{a}_{+} \hat{b}  + \epsilon^{2} - M^{2} ) \Phi_{2} = 0 \; ,
$$
$$
(- \hat{b}_{-} \hat{a} - \hat{a}_{+} \hat{b}
 - M^{2} ) \Phi_{0} + i \epsilon \; f =0\; ,
$$
$$
- \hat{b}_{-} \hat{a}  {f + g \over  2}   + (\epsilon^{2} - M^{2})  {f + g \over  2}    +
\hat{b}_{-} \hat{a} {f - g \over  2}  + i \epsilon \hat{b}_{-} \hat{a} \Phi_{0} = 0 \; ,
$$
$$
-  \hat{a}_{+} \hat{b}  {f - g \over  2} + (\epsilon^{2} - M^{2}) {f - g \over  2}   +
\hat{a}_{+} \hat{b} {f + g \over  2}  + i \epsilon  \hat{a}_{+} \hat{b} \Phi_{0} = 0 \; .
\eqno(4.4)
$$

Summing and subtract ing equations 3 and , we get
$$
(- \hat{b}_{-} \hat{a} - \hat{a}_{+} \hat{b}  + \epsilon^{2} - M^{2} ) \Phi_{2} = 0 \; ,
$$
$$
(- \hat{b}_{-} \hat{a} - \hat{a}_{+} \hat{b}
 - M^{2} ) \Phi_{0} + i \epsilon \; f =0\; ,
$$
$$
( - \hat{b}_{-} \hat{a}   + \hat{a}_{+} \hat{b})  g    + (\epsilon^{2} - M^{2})  f
      + i \epsilon  ( \hat{b}_{-} \hat{a} + \hat{a}_{+} \hat{b} )\;  \Phi_{0} = 0 \; ,
$$
$$
( - \hat{b}_{-} \hat{a} - \hat{a}_{+} \hat{b})  g    + (\epsilon^{2} - M^{2})  g
    + i \epsilon ( \hat{b}_{-} \hat{a}- \hat{a}_{+} \hat{b})  \Phi_{0} = 0 \; ,
\eqno(4.5)
$$

\noindent
Taking into account identities
$$
- \hat{b}_{-} \hat{a} - \hat{a}_{+} \hat{b}= \Delta_{2},\qquad
- \hat{b}_{-} \hat{a}   + \hat{a}_{+} \hat{b}=  2B
\eqno(4.6)
$$

\noindent
eqs. (4.5) reduce to the form
$$
(\Delta_{2}  + \epsilon^{2} - M^{2} ) \; \Phi_{2} = 0 \; ,
\eqno(4.7)
$$
$$
( \Delta_{2}  - M^{2} ) \; \Phi_{0} + i \epsilon \; f =0\; ,
$$
$$
2B  \;  g    + (\epsilon^{2} - M^{2})  f
      -  i \epsilon  \Delta_{2} \;  \Phi_{0} = 0 \; ,
$$
$$
\Delta_{2}  g    + (\epsilon^{2} - M^{2})  g
    - 2i \epsilon B \;   \Phi_{0} = 0 \; ,
\eqno(4.8)
$$

From the second equation, with the use of expression for $\Delta_{2} \Phi_{0}$ according to the first equation,
we  derive linear relation between three functions
$$
2B\; g -M^{2} f - i\epsilon M^{2} \Phi_{0} = 0 \; .
\eqno(4.9)
$$

\noindent With the help of (4.9), let us exclude  $f$
$$
 f = {2B \over M^{2}} \; g  - i\epsilon  \Phi_{0}
$$

\noindent from equations  2 and  3:
$$
(\Delta_{2}     + \epsilon^{2} - M^{2}) \;  g
    = 2i \epsilon B \;   \Phi_{0} \; ,
    $$
    $$
(\Delta_{2}     + \epsilon^{2} - M^{2}) \; \Phi_{0}  = - {2i\epsilon B  \over  M^{2}} \; g  .
\eqno(4.10)
$$

With notation  $\gamma = \epsilon^{2} / M^{2}$, the system can be written in a matrix form
$$
(\Delta_{2}     + \epsilon^{2} - M^{2})
\left | \begin{array}{r}
g  \\
\epsilon \; \Phi_{0}
\end{array} \right | =
\left | \begin{array}{rr}
0 & 2iB  \\
-2iB \gamma  & 0
\end{array} \right |
\left | \begin{array}{r}
g  \\
\epsilon \; \Phi_{0}
\end{array} \right |
\eqno(4.11)
$$

\noindent
or symbolically as
$$
\Delta f = A f \qquad  \Delta f ' = S AS^{-1} \;f'\;, \qquad f ' = S f \; .
$$

It remains to find a transformation reducing the matrix $A$ to a diagonal form
$$
S AS^{-1} = \left | \begin{array}{cc}
\lambda_{1} & 0 \\
0 & \lambda_{2}
\end{array} \right | , \qquad S = \left | \begin{array}{cc}
a & d \\
c & b
\end{array} \right | \; ;
$$

\noindent  the problem is equivalent to the linear system
$$
-\lambda_{1}  \; a - 2i\gamma B \; d  = 0 \; ,\qquad
2iB \; a  - \lambda_{1} \; d = 0 \; ;
$$
$$
-\lambda_{2}  \; c - 2i\gamma B \; b  = 0 \; ,
\qquad
2iB \; c  - \lambda_{2} \; b = 0 \; .
$$

\noindent Its solutions can be chosen in the form
$$
  \lambda_{1} =  + {2\epsilon B \over M} \;, \qquad \lambda_{2} =
-  {2\epsilon B \over M} \;, \; S= \left | \begin{array}{cc}
\epsilon & +iM \\
\epsilon & -iM
\end{array} \right |,
$$
$$
S^{-1}  = {1 \over -2i\epsilon M}
\left | \begin{array}{cc}
-i M & -i M \\
-\epsilon & \epsilon
\end{array} \right | .
\eqno(4.12)
$$

New (primed) function satisfy the following equations
$$
1) \qquad \left (\Delta_{2}     + \epsilon^{2} - M^{2} - {2\epsilon B \over M} \right  ) \; g'  = 0 \; ,
\eqno(4.13a)
$$
$$
2) \qquad \left (\Delta_{2}     + \epsilon^{2} - M^{2} + {2\epsilon B \over M}  \right  ) \;  \Phi_{0}'  = 0   \; .
\eqno(4.13b)
$$

\noindent They independent from each other. There exist two linearly independent ones
$$
1) \qquad  g' \neq 0 , \qquad \Phi'_{0}= 0 \; , \qquad 2) \qquad
g' =0 , \qquad \Phi'_{0} \neq  0 \; . \eqno(4.14)
$$

The initial functions for these two cases assume respectively the form
$$
1)  \qquad  g = {1 \over 2\epsilon } g'  \; ,\qquad \epsilon
\Phi_{0} = {1 \over  2i M} g' \; ;
$$
$$
2) \qquad  g =  {1 \over 2i\epsilon} \epsilon \Phi'_{0} \; ,\qquad
\epsilon \Phi_{0} =  - {1  \over 2iM} \epsilon\Phi_{0}' \; .
$$

In each  case, eqs. (4.10)   have the same form
coinciding with $(4.13a)$ and  $(4.13b)$ respectively.
To obtain explicit solutions for these differential equation, we need not any additional calculations,
instead it suffices to perform simple formal  changes as pointed below
$$
2\,\epsilon \, M \qquad \Longrightarrow \qquad
\left \{ \begin{array}{l}
(\epsilon^{2} - M^{2}  - {2\epsilon B \over M} )  \\
(\epsilon^{2} - M^{2} ) \\
(\epsilon^{2} - M^{2} + {2\epsilon B \over M} )
\end{array} \right.
\eqno(4.15)
$$

\section{ Acknowledgment }

 Authors are  grateful  to  participants of
seminar of Laboratory of theoretical physics,
 National Academy of Sciences of Belarus, for stimulating discussion and  advice.


\begin{thebibliography}{xxx}




\bibitem{1} Rabi I.I.
 Das freie Electron  in Homogenen Magnetfeld nach der  Diraschen Theorie.
// Z. Phys.    {\bf 49},  507 -- 511 (1928).



\bibitem{2}  Landau  L., Diamagnetismus der Metalle,
 Ztshr. Phys.    {\bf 64},   629--637 (1930).




\bibitem{3}
 Plesset M.S.
 Relativistic  wave  mechanics of the electron  deflected  by magnetic field.
// Phys.Rev.   {\bf  12}, 1728 -- 1731 (1931).


\bibitem{Landau-Lifshitz}
L.D. Landau,  E.M. Lifshitz. Quantum mechanics. Addison Wesley,
Reading, Mass., 1958.







\bibitem{Comtet-1985}
A. Comtet, P.J. Houston. Effective action on the hyperbolic plane
in a constant external field. J. Math. Phys.  1985. Vol. 26, No.
1. P. 185 -- 191





\bibitem{Comtet-1987}
Alain Comtet. On the Landau levels on the hyperbolic plane. Annals
of Physics. 1987. Vol.  Volume 173.  P.  185 -- 209.


\bibitem{Aoki-1987}
H. Aoki. Quantized Hall Effect.
 Rep. Progr. Phys. 1987. Vol. 50. P.  655 -- 730.


\bibitem{Groshe-1988}
C. Groshe. Path integral on the Poincar\'e uper half
 plane  with a magnetic field and for the Morse potential.
Ann. Phys. (N.Y.),  1988.  Vol. 187. P. 110 -- 134.


\bibitem{Klauder-Onofri}
J.R. Klauder,  E. Onofri. Landau Levels and Geometric
Quantization.
 Int. J. Mod. Phys.  1989. Vol. A4.  P. 3939 -- 3949.





\bibitem{Avron-Pnueli-1990}
J.E. Avron,  A. Pnueli.
 Landau Hamiltonians on Symmetric Spaces.
 Pages 96 -- 117 in:   Ideas and methods in mathematical analysis, stochastics, and
   applications.
 Vol. II. S. Alverio et al., eds. (Cambridge Univ. Press, Cambridge, 1990).







\bibitem{Plyushchay-1991(1)}
M.S. Plyushchay. The Model of relativistic particle with torsion.
Nucl. Phys. 1991.  Vol. B362. P. 54 -- 72.

\bibitem{Plyushchay-1991(2)}
M.S. Plyushchay. Relativistic particle with torsion, Majorana
equation and fractional spin. Phys. Lett. 1991. Vol. B262. P. 71
-- 78.


\bibitem{Dunne-1992}
G.V. Dunne. Hilbert Space for Charged Particles in Perpendicular
Magnetic Fields.
 Ann. Phys. (N.Y.)  1992. Vol.  215. P.  233 -- 263.

\bibitem{Plyushchay-1995}
M.S. Plyushchay. Relativistic particle with torsion and charged
particle in a constant electromagnetic field: Identity of
evolution. Mod. Phys. Lett. 1995.  Vol. A10. P. 1463 -- 1469;
hep-th/9309147.


\bibitem{Alimohammadi-Shafei Deh Abad-1996}
M. Alimohammadi,  A.Shafei Deh Abad. Quantum group symmetry of the
quantum Hall effect on the non-flat surfaces. J. Phys. 1996.  Vol.
A29. P.  559.



\bibitem{Alimohammadi-Mohseni Sadjadi-1996}
M. Alimohammadi, H.Mohseni Sadjadi
 Laughlin states on the Poincare  half-plane and their quantum group symmetry,
  Jour. Phys.  1996. Vol. A29. P.  5551



\bibitem{Onofri-2001}
E. Onofri. Landau Levels on a torus. Int. J. Theoret. Phys., 2001,
Vol. 40, no 2, P. 537 -- 549; arXiv:quant-ph/0007055v1 18 Jul 2000



\bibitem{Negro et al-2001}
J. Negro, M.A. del Olmo, A. Rodr\'{i}guez-Marco. Landau quantum
systems: an approach based on symmetry. arXiv:quantum-ph/0110152.



\bibitem{Gamboa et al-2001}
J. Gamboa, M. Loewe, F. Mendez, J. C. Rojas
  The Landau problem and noncommutative quantum mechanics.
  Mod. Phys. Lett. A.  2001.  Vol. 16. P.   2075 -- 2078.


\bibitem{Klishevich-Plyushchay-2002}
S.M. Klishevich, M.S. Plyushchay. Nonlinear holomorphic
supersymmetry on Riemann surfaces. Nucl. Phys. 2002.  Vol. B 640.
P. 481 -- 503; hep-th/0202077.


\bibitem{Drukker et  al-2004}
 N. Drukker, B. Fiol, J. Sim\'{o}n.
 G\"{o}del-type
Universes and the Landau problem. hep-th/0309199. Journal of
Cosmology and Astroparticle Physics (JCAP) 0410 (2004)  Paper 012



\bibitem{Ghanmi-Intissar-2004}
A. Ghanmi, A. Intissar. Magnetic Laplacians of fifferentila forms
of the hyperbolic disk and Landau levels. African Journal Of
Mathematical Physics. 2004.  Vol.  1.  P.  21 -- 28.




\bibitem{Correa-Jakubsky-Plyushchay-2009}
F. Correa, V. Jakubsky, M.S. Plyushchay. Aharonov-Bohm effect on
AdS(2) and nonlinear supersymmetry of reflectionless Poschl-Teller
system. Annals Phys. 2009. Vol. 324. P. 1078 -- 1094,2009;
arXiv:0809.2854.

\bibitem{Alvarez-Cortes-Horvathy-Plyushchay-2009}
P.D. Alvarez, J.L. Cortes, P.A. Horvathy, M.S. Plyushchay.
  Super-extended noncommutative Landau problem and conformal symmetry.
JHEP. 2009.  0903:034; arXiv:0901.1021.



\bibitem{4}
  A.A. Bogush,  V.M.  Red'kov,   G.G. Krylov.
Schr\"{o}dinger particle in magnetic and electric fields in
Lobachevsky and Riemann spaces. // Nonlinear Phenomena in Complex
Systems.  {\bf  11},  no 4,  403 -- 416 (2008).


\bibitem{5}
A.A. Bogush,  G.G. Krylov,  E.M. Ovsiyuk, V.M. Red'kov. Maxwell
electrodynamics in complex form, solutions with cylindric symmetry
in the Riemann space. Doklady Natsionalnoi Akademii Nauk Belarusi.
{\bf 33},   52 -- 58 (2009).


\bibitem{6}
A.A. Bogush, V.M. Red'kov, G.G. Krylov. Quantum-mechanical
particle in a uniform magnetic field in spherical space $S_{3}$.
 Proceedings of the National Academy of Sciences of Belarus. Ser. fiz.-mat. {\bf 2}, 57 -- 63 (2009).




\bibitem{7}
 V.V. Kudryashov, Yu.A. Kurochkin, E.M. Ovsiyuk, V.M.
Red'kov. Motion caused by magnetic field in Lobachevsky space. AIP
Conference Proceedings. Vol. 1205, P. 120 -- 126 (2010); Eds. Remo
Ruffini and Gregory Vereshchagin. The sun, the stars, the Uiverse
and General relativity. International Conference in Honor of Ya.B.
Zeldovich. April 20-23, 2009, Minsk.



\bibitem{8}
V.V. Kudryashov, Yu.A. Kurochkin, E.M. Ovsiyuk, V.M.  Red'kov.
Motion of a particle in magnetic field in the Lobachevsky space.
Doklady Natsionalnoi Akademii Nauk Belarusi.  {\bf 53},  50--53
(2009).


\bibitem{9}
V.V. Kudryashov, Yu.A. Kurochkin, E.M. Ovsiyuk, V.M. Red'kov.
Classical Particle in Presence of Magnetic Field, Hyperbolic
Lobachevsky and Spherical Riemann Models.
 SIGMA {\bf  6}, 004, 34 pages (2010).

\bibitem{Book-1}
V.M.  Red'kov. Fields in Riemannian space  and the Lorentz group.
Publishing House "Belarusian Science", Minsk, 2009 (in Russian).






\bibitem{Olevsky}
M.N. Olevsky. Three-orthogonal  coordinate systems in spaces of
constant curvature, in which equation  $\Delta_{2}U + \lambda U=0$
permits the  full separation of variables. Mathematical
collection.  1950. Vol.  27. P. 379 -- 426.



\bibitem{Bateman}
H. Bateman, A. Erd\'elyi.   Higher transcendental functions. Vol.
I. (New York, McGraw-Hill) 1953.



\bibitem{Book-5}
  V.M. Red'kov, E.M. Ovsiyuk.
 Quantum mechanics in spaces of  constant curvature.
Nova Science Publishers. Inc. 2011.  (in press)


\end{thebibliography}
\end{document}